\documentclass{PoS}

\title{Stochastic acceleration in accreting black holes}
\ShortTitle{Stochastic acceleration in accreting black holes}

\author{\speaker{Alexandre Marcowith}\\
        Laboratoire Univers et Particules de Montpellier, Universit\'e Montpellier II/CNRSFrance\\
        E-mail: \email{Alexandre.Marcowith@univ-montp2.fr}}

\author{Renaud Belmont \thanks{Corresponding author}\\
        Institut de Recherche en Astrophysique et Plan\'etologie de Toulouse, Universit\'e Paul Sabatier/CNRS, France\\
       E-mail: \email{Renaud.Belmont@irap.omp.eu}}

\author{Julien Malzac\\
        Institut de Recherche en Astrophysique et Plan\'etologie de Toulouse, Universit\'e Paul Sabatier/CNRS, France\\
        E-mail: \email{Julien.Malzac@irap.omp.eu}}

\abstract{We present an extension of the BELM code (Belmont et al 2008) to investigate the microphysics of particle acceleration in black holes 
accretion disc corona. The updated version of the code accounts for the dynamics of resonant slab waves as well as their interaction with both
leptons or protons. It is found that the proton temperature is an important regulating effect of the stochastic particle acceleration process in accretion disk corona. We present a preliminary fit of the high soft spectral state of Cygnus X-1.}

\FullConference{"An INTEGRAL view of the high-energy sky (the first 10 years)"
9th INTEGRAL Workshop and celebration of the 10th anniversary of the launch,\\
		October 15-19, 2012\\
		Biblioth\`eque Nationale de France, Paris, France}

\begin{document}

\section{Introduction}
Black holes in X-ray binary systems are characterized by two main spectral states (see \cite{Malzac09}). At luminosities close to the Eddington limit the spectrum is dominated by a thermal component at keV energies and by a soft non-thermal tail  with a photon index $\Gamma \sim 2.3-3.5$ extending up to MeV energies. In this state the source is bright in soft X-rays and faint in the hard X-rays; it is called the High Soft State (hereafter HSS). At low luminosities (about 1\% of the Eddington luminosity) the soft X-ray component is faint and the spectrum is dominated by a hard power-law tail with a photon index $\Gamma \sim 1.5-1.9$. In this state the source is faint in soft X-rays and strong in hard X-rays; it is called the Low Hard State (hereafter LHS). These spectral states are usually explained by an hybrid thermal and non-thermal corona model. The LHS state is usually well fitted by a model including multiple Compton up-scatterings of soft photons by a Maxwellian distribution of electrons  in a hot (kT $\sim$ 50-100 keV) plasma of Thomson optical depth of the order of unity. The HSS is usually interpreted as inverse Compton up-scattering of soft disk photons (UV, soft X) by an hybrid thermal/non-thermal distribution in a hot coronal plasma. \\ 
Now the origin of the electron energization is still widely debated. One of the favoured process is Coulomb heating by a hot population of ions \cite{Nayakshin98}, but magnetic reconnection above the accretion disk \cite{Galeev+79} or turbulent heating by the interplay of magneto-hydrodynamic (MHD) turbulence \cite{Quataert99} are likely to operate the magnetic field effects are included. This work considers the latter process as the main source of electron (and/or positron) acceleration process and investigate the possibility to build-up non-thermal particle distribution. We propose a microphysical process describing quantitatively the transition between the LHS and HSS spectral states. 

\section{Physical modelling}
The hot corona is approximated as a sphere of radius R inside which photons, energetic leptons (electrons and possibly positrons) and magnetized plasma fluctuations are in a close interaction. We also account for the presence of a thermal population of protons with a fixed temperature which will contribute to the electron Coulomb heating.  We consider a simplified model in which magnetic fluctuations are treated as slab MHD modes propagating forward and backward with respect to the mean magnetic field. The waves are present at all wavenumbers as part of a turbulent cascade. The turbulent energy is injected at large scales $L_{inj} =R$ and proceeds towards smaller scales through non-linear wave-wave interactions down to scales where the Landau damping by the different particle species takes over. Waves are chosen in two  circular polarization states left (L) and right (R) depending on the clockwise or anti-clockwise electric field rotation direction. Waves and particles interact only via gyroresonance (transit-time damping will be accounted elsewhere). A wave of frequency $\omega$ at a particle wave-number k and a particle interact through gyro-resonance described by the following condition \cite{Steinacker92}
\begin{equation}
\label{Eq:Res}
\omega - kv_{\parallel} = n {\Omega_c \over \gamma} \ , \ \rm{n=\pm1} \ ,
\end{equation}
where $\Omega_c$ and $\gamma$ are the cyclotron frequency and the particle Lorentz factor respectively. Electrons, positrons and protons can interact with L and R modes in the lab rest frame. \\

\begin{figure} 
\includegraphics[width=.7\textwidth]{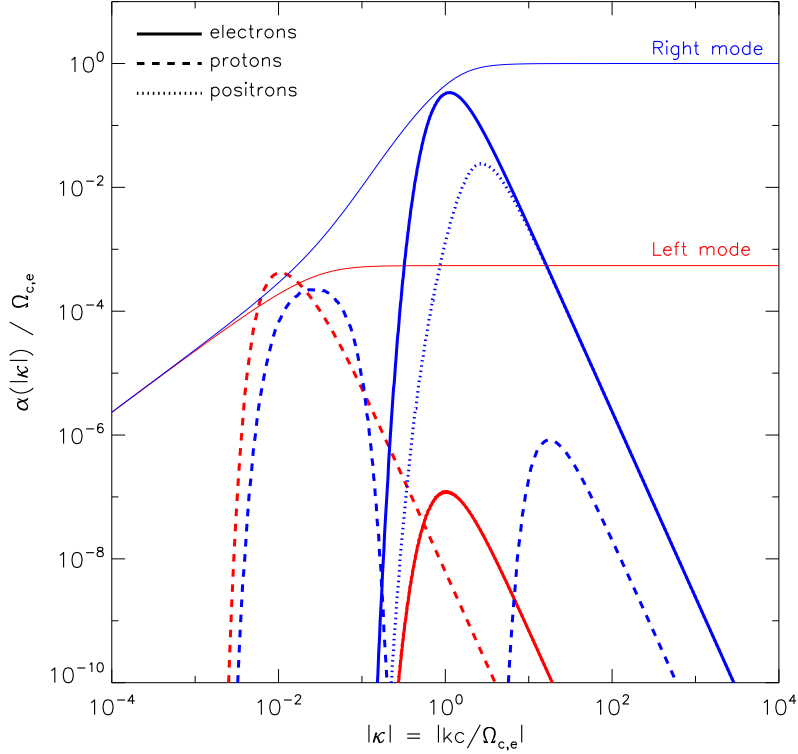} 
\caption{L (red) and R (blue) modes absorption rates (normalized to the electron cyclotron frequency) with respect to the normalized wavenumber by a population of thermal electrons (continuous line) and thermal protons (dashed line) respectively. Also displayed are the dispersion relations of the two modes in light continuous lines. R mode absorption rate by positrons is shown in dotted blue line, the absorption rate of L modes is superimposed with the electron curve.} 
\label{F:Damp} 
\end{figure}

\begin{figure} 
\includegraphics[width=.7\textwidth]{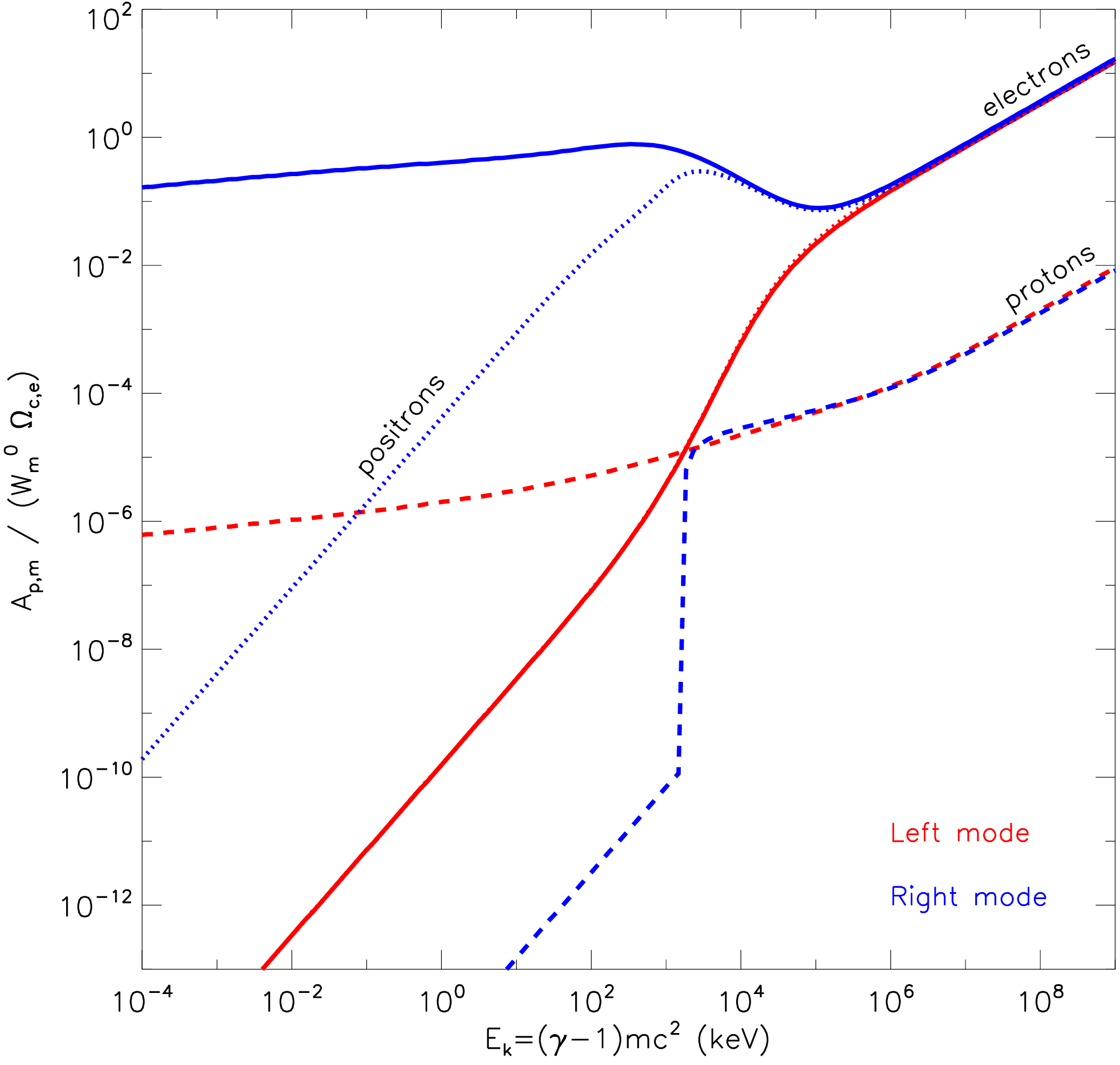} 
\caption{Electron (continuous line), positron (dotted line) and proton (dashed line) acceleration rate wrt the particle's kinetic energy by different wave species (L modes in red, R modes in blue).} 
\label{F:Acc} 
\end{figure}

The wave, photon and particles dynamics are treated by solving {\it three} coupled kinetic equations (see section \ref{S:Num}). The wave absorption coefficient 
$\alpha(k)$ depends on a kernel describing the wave-particle resonance absorption effect modulated by the particle distribution $N(p)$. The particle acceleration $A(p)$ rate is similarly calculated with the help of the same kernel modulated by the wave distribution $W(k)$. This formalism naturally leads to energy conservation as $\int dk \alpha(k) = \int dp A(p)$. To illustrate the respective impacts of waves and particles we first display in figure \ref{F:Damp} the absorption of L and R waves by a population of thermal electrons of temperature of 100 keV and a population of thermal protons of temperature of 1 MeV respectively. As can be seen protons mostly absorb L modes whereas electrons (to a lesser degree positrons) and protons compete in the absorption process of R modes. 
In figure \ref{F:Acc} we display the particle acceleration rate by the two wave species with respect to the particle kinetic energy . 

\section{Numerical modelling}
\label{S:Num}
The waves non-linear dynamics are followed using a diffusion term in a kinetic equation as described in \cite{Zhou90}. The electron and photon dynamics are followed solving a set of coupled non-linear kinetic equations as described in \cite{Belmont08}. This work is an updated modeling with respect to \cite{Belmont08} using an accurate calculation of particle acceleration and wave damping processes. We account at each timestep for the resonances between the particles and the different wave species (Belmont et al 2013, in prep) as described in Eq~\ref{Eq:Res}. The code has several inputs: the Thomson optical depth $\tau$, the magnetic compactness $l_B= (\sigma_T/8\pi m_e c^2) R B^2$, the disk soft photon compactness $l_s=(\sigma_T/m_e c^2) L_s/R$, where $L_s$ is the disk luminosity, the disk temperature $T_{bb}$, the proton temperature $T_p$ and the turbulence compactness $l_t=  (\sigma_T/8\pi m_e c^2) R \delta B^2$ and the turbulence index $q$.

\subsection{Case study}
\begin{figure} 
\includegraphics[width=.75\textwidth]{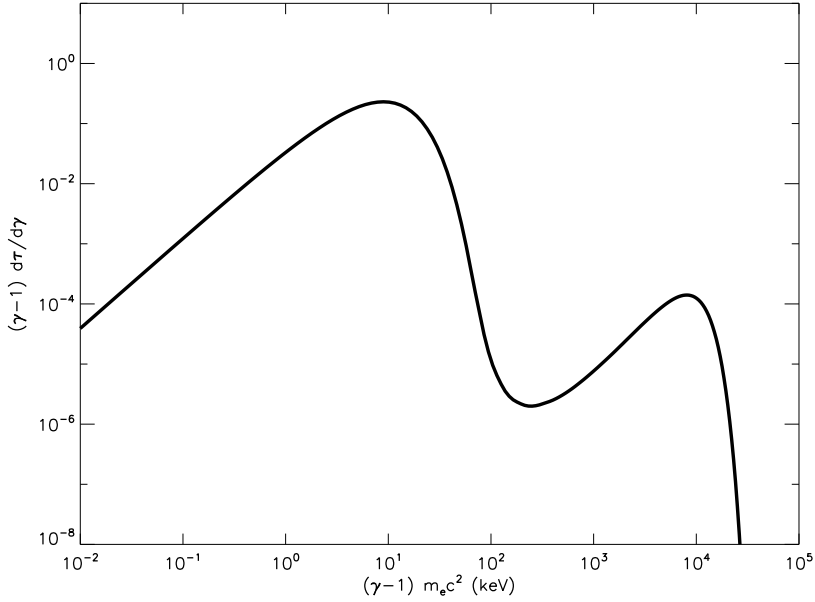} 
\caption{Electron distribution function wrt particle kinetic energy produced in the fiducial example.} 
\label{F:Part} 
\end{figure}

We illustrate the outputs obtained from the updated code using the following fudicial example with $R=5 10^7 \ \rm{cm}$, $\tau=0.5$, $l_s=5 10^{-3}$, $\ell_B=5 10^{-2}$, $\beta_p=0.1$ (plasma beta parameter), $T_{bb}=0.38$, $q=5/3$, $l_{t}=0.025$. The figure \ref{F:Part} provides the outcoming particle spectrum obtained after the turbulent spectrum has reached the damping scales. One can see a dual electron population (one thermal and one supra-thermal). The accelerated electrons do not produced a non-thermal power-law because of the important radiative losses that produce a pile-up effect. 

\subsection{The proton switch effect}
The protons are very efficient in absorbing L and R modes. It happens that if the proton temperature exceeds a certain threshold most of the turbulent energy is used in heating the protons rather than accelerating the electrons. This threshold corresponds to a plasma parameter of the order of one and can be written for typical numbers adapted for X-ray binaries as: 
\begin{equation}
T_{thr} \simeq [1 \ \rm{MeV}] \times \tau^{-1} \times \left({B \over 10^6 \ \rm{Gauss}}\right)^2 \times \left({R \over 10^8 \ \rm{cm}}\right) \ .
\end{equation}
Hence if $T_p \le T_{thr}$ the electron distribution harbours a suprathermal component likewise in figure \ref{F:Part}, whereas if $T_p> T_{thr}$ the electron distribution is reduced to a thermal component: no strong acceleration does occur. The proton switch effect is likely a natural regulation process in these plasmas.

\section{Spectral states of Cygnus X-1}
\begin{figure} 
\includegraphics[angle=-90, width=.8\textwidth]{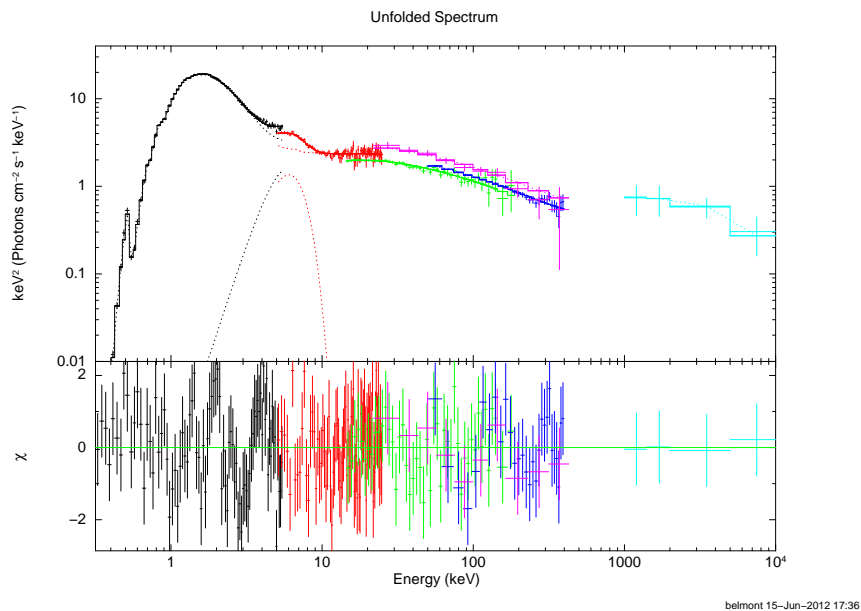} 
\caption{The HSS of Cygnus X-1 fitted using the updated BELM code. The reduced $\chi^2$ is 231/236. Data have been taken from \cite{McConnell02}.} 
\label{F:Cyg} 
\end{figure}

We provide here a {\it preliminary} fit of the source Cygnus X-1 in the high soft state in figure \ref{F:Cyg}. The parameters are the following: $\tau=0.046$, $l_s=0.011$, $l_B=0.19$, $\beta_p=0.33$, $l_s/l_t=0.2$. In that case the proton temperature is low enough to permit R modes to cascade down to energies where they interact resonantly with electrons. A specific work including fits of Cygnus X-1 in the LHS and the transition between the two states will be presented elsewhere. 

\section{Conclusions}
The BELM code \cite{Belmont08} has been updated to account for the production and the effect of turbulent magnetic fluctuations over particle acceleration in black hole binary systems. The turbulent spectrum consists of slab modes with state of polarization either left-handed or right-handed. Protons and leptons can have resonant interactions with each type of modes. It is found that the proton temperature is an important regulating effect of the stochastic particle acceleration process in accretion disk corona as the entire  R mode cascade can get absorbed before being able to interact with electrons. In that case the electron distribution is reduced to be thermal. Alternatively if proton temperature is small enough hence suprathermal electrons are produced and they can comptonize both synchrotron and disk radiation. A preliminary fit shows that the high soft spectral state of Cygnus X-1  can be well reproduced. \\

The authors thank A.Zdziarski for providing them with the Cygnus X-1 data taken from \cite{McConnell02}.


\begin{thebibliography}{99}
\bibitem{Malzac09} Malzac J.\& Belmont R., 2009, MNRAS, 392, 570
\bibitem{Nayakshin98} Nayakshin S., Melia F., 1998, ApJS, 114, 269
\bibitem{Galeev+79} Galeev A. A., Rosner R., Vaiana G. S., 1979, ApJ, 229, 318 
\bibitem{Quataert99} Quataert E., Gruzinov A., 1999, ApJ, 520, 248
\bibitem{Steinacker92} Steinacker J. \& Miller, J.A., 1992, ApJ, 393, 764
\bibitem{Zhou90} Zhou Y. \& Matthaeus W.H., 1990, JGR, 95, 14881
\bibitem{Belmont08} Belmont, R., Malzac, J. \& Marcowith, A., 2008, A\&A, 491, 617
\bibitem{McConnell02} McConnell  M.L. et al, 2002, ApJ, 572, 984
\end{thebibliography}
\end{document}